\documentclass[prb,twocolumn,showpacs,amsmath,amssymb,superscriptaddress]{revtex4-1}
\usepackage{epsfig,epstopdf} 
\usepackage{bbm}
\usepackage{mathrsfs}
\usepackage{epsfig}
\usepackage{graphicx}
\usepackage{amsfonts}
\usepackage[figuresright]{rotating}
\usepackage{amssymb}
\usepackage{amsmath}
\usepackage{dcolumn}
\usepackage{bm}
\usepackage{braket}
\usepackage{comment}
\usepackage{mathtools}
\usepackage{xcolor}
\usepackage{multirow}
\usepackage{setspace}
\usepackage{multirow}
\usepackage{hyperref}

\begin{document}
\title{Interacting topological mirror excitonic insulator in one dimension}
\author{Lun-Hui Hu}
\email{l1hu@physics.ucsd.edu}
\affiliation{Department of Physics, University of California, San Diego, California 92093, USA}
\affiliation{Kavli Institute of Theoretical Sciences, University of Chinese Academy of Sciences, Beijing, 100049, China}
\author{Rui-Xing Zhang}
\email{ruixing@umd.edu}
\affiliation{Condensed Matter Theory Center and Joint Quantum Institute, Department of Physics, University of Maryland, College Park, Maryland 20742-4111, USA}
\author{Fu-Chun Zhang}
\affiliation{Kavli Institute of Theoretical Sciences, University of Chinese Academy of Sciences, Beijing, 100049, China}
\affiliation{CAS Center for Excellence in Topological Quantum Computation, University of Chinese Academy of Sciences, Beijing, 100049, China}
\affiliation{Collaborative Innovation Center of Advanced Microstructures, Nanjing University, Nanjing 210093, China}
\author{Congjun Wu}
\affiliation{Department of Physics, University of California, San Diego, California 92093, USA}

\begin{abstract}
We introduce the topological mirror excitonic insulator as a new type of interacting topological crystalline phase in one dimension. Its mirror-symmetry-protected topological properties are driven by exciton physics, 
and it manifests in the quantized bulk polarization and half-charge modes on the boundary. 
And the bosonization analysis is performed to demonstrate its robustness against strong correlation effects in one dimension. 
Besides, we also show that Rashba nanowires and Dirac semimetal nanowires could provide ideal experimental platforms to realize this new topological mirror excitonic insulating state. 
Its experimental consequences, such as quantized tunneling conductance in the tunneling measurement, are also discussed. 
\end{abstract}  

\maketitle

\section{Introduction}
The conceptual revolution of topological physics \cite{hasan2010,qi2011,hasan2011,konig2008,maciejko2011,zhang2013,wehling2014,yan2012,ren2016,galitski2013,goldman2014,bansil2016,chiu2016,cooper2018} in solids has changed our way of classifying phases of matters, 
which has made a great impact on the experimental discoveries of the topological insulators (TI) \cite{konig2007,knez2012,chen2009,wu2018}. 
By definition, a time-reversal-invariant TI is featured by the nontrivial $\mathbb{Z}_2$ band topology and the resulting gapless boundary modes,
both of which are protected by the time-reversal symmetry \cite{bernevig2013,franz2013,shen2012}.
This symmetry protection \cite{chiu2016}, only upon which the $\mathbb{Z}_2$ topological invariant is well-defined, distinguishes TI from the earlier examples such as the integer/fractional quantum Hall systems. 
And it leads to the concept of the symmetry-protected topological (SPT) state \cite{chen2012}. 
The framework of symmetry-protected band topology has been successfully extended from the time-reversal symmetry to the crystalline topological systems protected by space-group symmetries \cite{fu2011,fang2012,slager2013,liu2014,wang2012,armitage2018}, magnetic-group symmetries \cite{mong2010,fang2013,zhang2015} and space-time symmetries \cite{xu2018,morimoto2017}. 
Remarkable progresses have been made along this direction, which has lead to the theoretical predictions and experimental discoveries of the crystalline topological phase in SnTe \cite{hsieh2012}, Pb$_{1-x}$Sn$_x$Se \cite{dziawa2012}, KHgSb \cite{wang2016,ma2017}, and recently in MnBi$_{2n}$Te$_{3n+1}$ \cite{zhang2019topological,otrokov2018prediction,li2019intrinsic,gong2019experimental,li2019dirac,hao2019gapless,chen2019topological,zhang2019m}.
  
Meanwhile, there have been great interests in exploring the fate of SPT phases under strong electron correlations \cite{hohenadler2013,senthil2015,rachel2018}. 
For example, interaction effects could (i) fundamentally change the topology by breaking symmetries spontaneously \cite{wu2006,xu2006,rachel2010,zheng2011}, or alter the topological classification of SPT states \cite{fidkowski2010,fidkowski2011,ryu2012,qi2013,yao2013}
, and (ii) enable new topological phases which do not exist in the free fermion limit \cite{wen2010,rachel2010b,pesin2010,neupert2011,sun2011,tang2011,budich2012,wang2012c,cocks2012,gu2014,wang2014b,you2014,bi2017}. 
However, little is known about how to realize an interacting SPT state in experiments \cite{dzero2010,ruegg2011,li2015,du2017,bi2017,zhang2017finger,xiao2018,de2019} as well as their detection, especially for those systems with a crystalline-symmetry protection. 

In this work, we introduce the topological mirror excitonic insulator (TMEI) as a new type of interacting crystalline topological state in one dimension (1D), and propose to experimentally realize this novel topological state in two physical systems, namely: the Rashba nanowires \cite{morales1998,spintronics2002,lauhon2002} and the Dirac semimetal nanowires \cite{li2015a,wang2016a,zhang2018b}. 
The TMEI is featured by a quantized bulk charge polarization and a mirror-protected boundary half-charge mode when interaction-induced excitonic order is formed. 
To fully incorporate the interaction effects in 1D, we construct an effective field theory description for the TMEI in the Rashba nanowire system by applying Abelian bosonization technique, which shows the robustness of the TMEI beyond the mean-field level. 
This allows us to map out the topological phase diagram and clarify the necessary condition for realizing the TMEI phase. 
Experimentally, we propose the quantized tunneling conductance as the smoking-gun signal for the TMEI, which clearly distinguishes the TMEI phase from other states.

\section{Model Hamiltonian}
The minimal system for 1D exciton physics consists of one electron band and one hole band \cite{su2008make}. 
Such two-channel systems can be realized in: 
(i) a double-wire setup with one $n$-type nanowire (electron doping) and a $p$-type nanowire (hole doping); 
(ii) a single quantum wire with two conducting channels that have opposite effective masses. 
In this Letter, we will focus on the TMEI physics in the double-nanowire setup and briefly discuss its realization in a single Dirac semimetal nanowire. 

\begin{figure}[!htbp]
\centering
\includegraphics[width=0.9\linewidth]{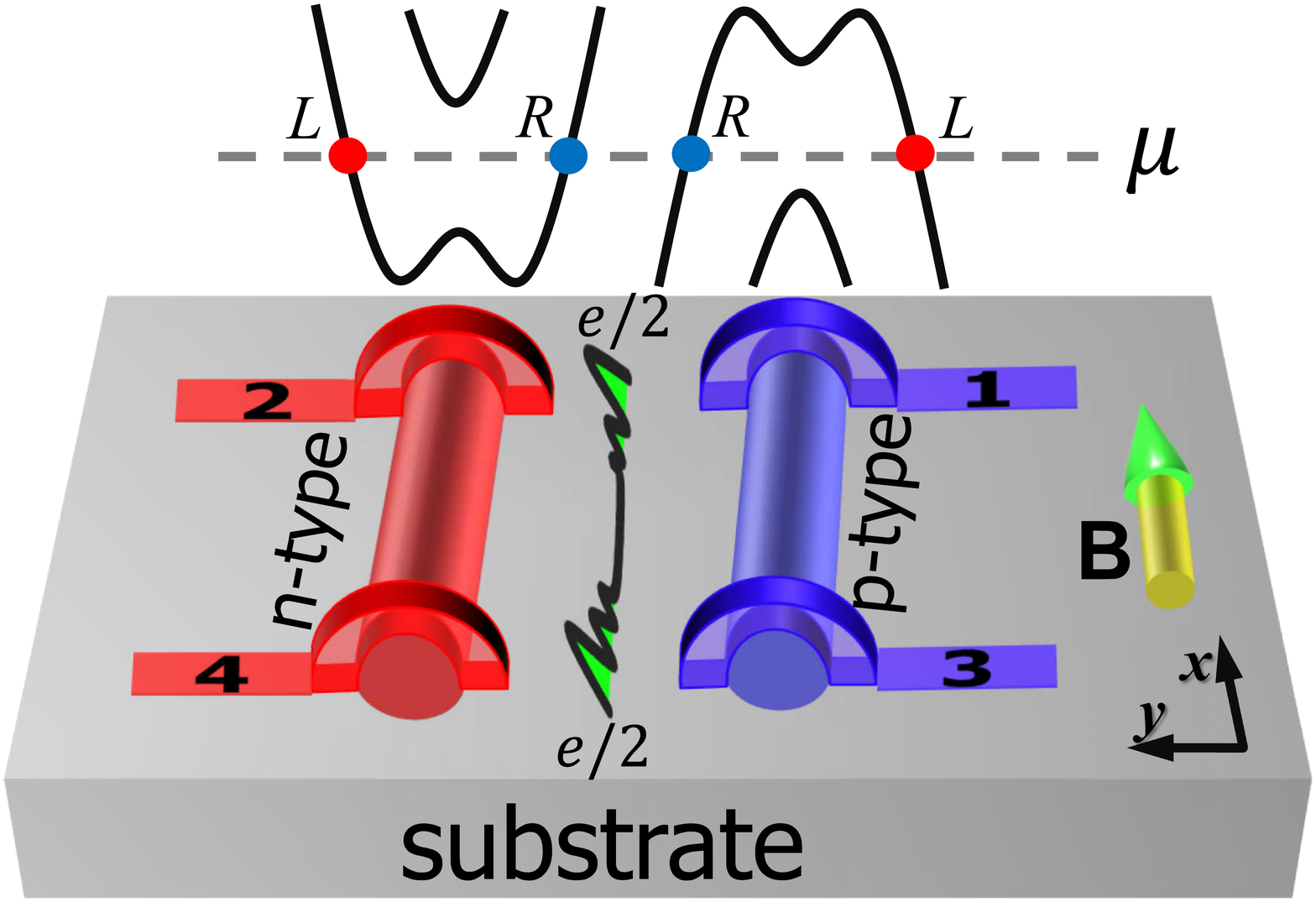}
\caption{\label{fig-sketch-nanowire} 
Schematics of two couple nanowires for topological mirror excitonic insulator. 	 
The left red wire is n-type and the right blue wire is p-type.
An in-plane magnetic field $\vec{B}$ is applied along the $\hat{x}$ direction. 	
The top panels show electron or hole dispersions for n-type (left panel) and p-type (right panel) wires in Eq.~\eqref{eq-ham0}, respectively, with only one left (L) moving and one right (R) moving  electron or hole at a chemical potential $\mu$ illustrated by a dashed line. 
A topological mirror excitonic insulator may be realized when the interaction between two nanowires is introduced, and half-charge modes are localized at the ends.
}	
\end{figure}

The double-nanowire system on an insulating substrate is illustrated in Fig.~\ref{fig-sketch-nanowire}. 
We consider a ${\bf k}\cdot{\bf p}$ Hamiltonian \cite{winkler2003} to describe the low-energy band structure of the double Rashba nanowires under a magnetic field along the wire direction $x$,
\begin{align}\label{eq-ham0}
\mathcal{H}_0 &= \epsilon(k_x) + (m_0 k_x^2 - \mu)\sigma_0 \tau_z + v k_x \sigma_y \tau_0 + h\sigma_x \tau_0,
\end{align}
in which Pauli matrices $\tau_i$ and $\sigma_i$ denote the wire and spin degrees of freedom, respectively; 
$\epsilon(k_x)=\delta \mu + \delta m k_x^2$;
$m_0$ is the inverse of the effective mass; 
$\mu$ is the chemical potential; 
$v$ characterizes the Rashba spin-orbit coupling; 
$h$ represents the Zeeman splitting energy induced by an applied magnetic field ${\bf B}$.
In the strained nanowires, $\delta m\to0$ can be achieved \cite{pryor2005}.
Here we assume that $m_0, v, h$ are all positive,
and there is a high tunneling barrier between the wires that suppresses the single electron tunneling between the wires. 
Nevertheless, the pair hopping processes still exist as a result of inter-wire interactions. 
When ${\bf B}$ is aligned along the wires ($\hat{x}$ direction), the Hamiltonian in Eq.~\eqref{eq-ham0} has a mirror symmetry $M_x = i\sigma_x\cdot P$, where $P$ maps $x$ to $-x$ 
\footnote{We note that $i\sigma_x\tau_z$ also acts like a mirror operation to Eq.~\eqref{eq-ham0}. 
However, we can explicitly break this accidental symmetry by turning on a small inter-wire tunneling term $t\tau_x$, which respects $M_x$.}.
In fact, $M_x$ is a lattice symmetry of the Rashba wires, which holds for both the lattice and continuum limits.

\section{Excitons and Topology}
To examine the band topology stemming from the exciton physics, 
we consider an inter-wire interaction,
\begin{align}
\mathcal{H}_{int} = \int dx \, \hat{n}_e(x)\hat{n}_h(x),
\end{align}
where $\hat{n}_i(x)=\sum_{\alpha=\uparrow,\downarrow}c^\dagger_{i\alpha}(x)c_{i\alpha}(x)$ is the density operator with $i=\{e,h\}$.
The excitonic orders \cite{hao2010,hao2011,budich2014,pikulin2014,chen2017} may introduce an energy gap in the double-wire system, in favor of the excitonic state energetically.
Based on the representations of $M_x$, there are two classes: 
\begin{align}
\label{Eq: mirror orders}
\begin{cases}
\;\text{Mirror-even orders}: \Delta_{\text{even}} = \sigma_{0(x)}\tau_{x(y)}, \\
\;\text{Mirror-odd orders}:  \Delta_{\text{odd}} = \sigma_{y(z)}\tau_{x(y)}.
\end{cases}
\end{align}
As a result, the mean-field Hamiltonian is given by,
\begin{align}
 \mathcal{H}_{MF}  = \mathcal{H}_0 +  H_{ec},
\end{align}
where $\mathcal{H}_0$ is the non-interacting Hamiltonian in Eq.~\eqref{eq-ham0} and $H_{ec}$ is for the excitonic orders in Eq.~\eqref{Eq: mirror orders}.
To show the gap opening by $H_{ec}$, we take it as a perturbation.
In the absence of $H_{ec}$, the eigenstates at $k_x=0$ are,
\begin{align}
\begin{split}
\vert h+\mu \rangle &= \left\lbrack 1, 1, 0, 0 \right\rbrack^{\text{T}}/\sqrt{2}\;\text{ and } \; m_x = +i, \\
\vert h-\mu \rangle &= \left\lbrack 0, 0, 1, 1 \right\rbrack^{\text{T}}/\sqrt{2}\;\text{ and } \; m_x = +i,  \\
\vert -h+\mu \rangle &= \left\lbrack 1, -1, 0, 0 \right\rbrack^{\text{T}}/\sqrt{2}\;\text{ and } \; m_x = -i, \\
\vert -h-\mu \rangle &= \left\lbrack 0, 0, 1, -1 \right\rbrack^{\text{T}}/\sqrt{2}\;\text{ and } \; m_x = -i.
\end{split}
\end{align}
where we assume $-h<\mu<h$, and $m_x$ is the eigenvalue of mirror symmetry $M_x$. 
By perturbation theory, we find that the mirror-even order parameters tends to lower the eigen-energy of the state $\vert h-\mu \rangle$, 
while it rises up the eigen-energy of the state $\vert -h+\mu \rangle$. 
The closing and reopening of the bulk gap indicates there is a topological phase transition.
More explicitly, the bulk dispersion is,
\begin{align}
E_{\pm}  = \delta\mu \pm \left\vert h \pm \sqrt{\mu^2+\Delta_0^2} \right\vert,
\end{align}
it shows gap closing when $h=\sqrt{\mu^2+\Delta_0^2}$, indicting that a topological phase transition can be tunned by the external magnetic field.
Moreover, the condition for the TMEI phase is,
\begin{align}
h>\sqrt{\mu^2+\Delta_0^2} \text{ with } \Delta_0\neq0, 
\end{align}
where $\Delta_0$ denotes the amplitude of the mirror-even excitonic orders. 
Given the TMEI condition, we note that the electron and hole bands are ``inverted" near the Fermi level to trigger the nontrivial band topology. 
This is shown in Fig.~\ref{fig-sketch-nanowire}, where we plot both bands for electron and hole nanowires in the $\Delta_0\to 0$ limit. 
As for the mirror-odd orders, they are trivial since they always increase the bulk gap.  
Namely, they generally spoil $M_x$ and do not lead to nontrivial topology in 1D. 

\begin{figure}[!htbp]
	\centering
	\includegraphics[width=\linewidth]{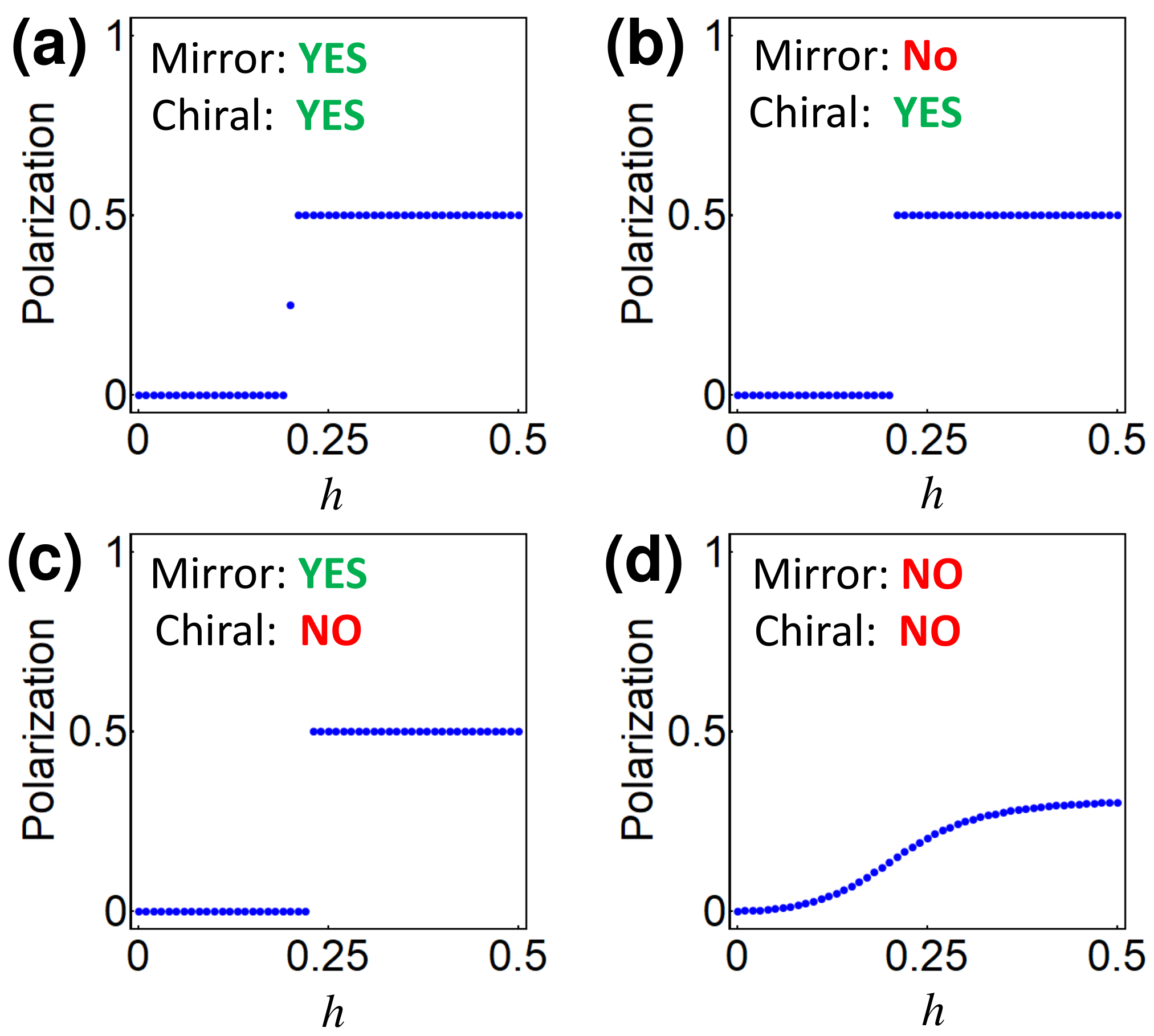}
	\caption{\label{sm-fig-polarization} 
		The Zak phase in different states.
		(a) Both mirror and chiral symmetries are preserved;
		(b) Mirror symmetry is broken but chiral symmetry is preserved;
		(c) Mirror symmetry is preserved but chiral symmetry is broken;
		(d) Both mirror and chiral symmetries are broken.
	}	
\end{figure}

With mirror-even orders, we can define a quantized bulk electric polarization ${\cal P}$ \cite{zak1989,resta1994,lau2018} to characterize the bulk topology,
\begin{align}
{\cal P} = \frac{i}{2\pi} \sum_{n} \oint dk_x \langle u_{n}(k_x) \vert \partial_{k_x} |u_{n} (k_x)\rangle ,
\end{align}
where the $|u_n(k_x)\rangle$ is the Bloch wave function with an occupied-band index $n$. 
Crucially, $M_x$ enforces ${\cal P}$ to be quantized to multiple of half integers (i.e. $0$ or $\frac{1}{2}$ since ${\cal P}$ is well-defined modulo $1$). 
${\cal P}$ can be numerically calculated within a mean field theory, and the results are shown in Fig.~\ref{sm-fig-polarization},
from which we find that only the system with the mirror symmetry can be topological.
And $\mathcal{P}=\frac{1}{2}$ for the TMEI phase, and $\mathcal{P}=0$ for the topological trivial phase. 

To drive the TMEI phase with ${\cal P}=\frac{1}{2}$ into a trivial phase (or a vacuum state) with ${\cal P}=0$, the system must undergo a bulk gap closing process, which manifests itself as a topological phase transition.
With an open boundary condition, one can easily check that the system hosts one localized half-charge mode at each end, similar to the case of the Su-Shrieffer-Heeger model.   
However, the chiral symmetry ${\cal C}$ is not often exact in our system, we emphasize that the mirror-enforced boundary half charge is more fundamental and robust than the ${\cal C}$-protected boundary zero mode in our TMEI phase

\section{Bosonization and Phase Diagram} 
Next, we establish a Luttinger liquid theory to show that the TMEI physics remains robust when interaction effects are considered, which is beyond the mean-field theory. 
Since there is a pair of counter-propagating modes in each wire [see Fig. \ref{fig-sketch-nanowire}],
we follow the standard mapping from the fermion fields $\Psi_{i,s}$ to the chiral boson fields $\chi_{i,s}$ by 
\begin{align}
\Psi_{i,s} = \frac{\eta_{i,s}}{\sqrt{4\pi a}} e^{ i s\sqrt{4\pi} \chi_{i,s}},
\end{align} 
where $a$ is the lattice constant and $\eta_{i,s}$ are Klein factors \cite{wu2003,wu2005}.
Here the wire index $i=\{e,h\}$ labels the $n$/$p$-type nanowire and $s=\pm$ denotes the right/left-moving fermion modes. 

It is convenient to define the dual boson fields as $\phi_{i} = \chi_{i,R} + \chi_{i,L}$ and $\theta_{i}=\chi_{i,R}-\chi_{i,L}$. 
Then the fermionic density operators are given by $\rho_{i,s}=\frac{1}{\sqrt{\pi}} \partial_x \chi_{i,s}$.
We first consider the intra-wire density-density interaction $g_1\rho_{i,L}\rho_{i,R}$, 
and the inter-wire density-density interaction $g_2(\rho_{e,R}\rho_{h,L}+\rho_{e,L}\rho_{h,R})$, which are known to renormalize the Fermi velocities and Luttinger parameters.
For simplicity, we assume that the bare Fermi velocities in these two nanowires are equal.
We denote $\phi_{\pm} = \left( \phi_{e} \pm \phi_{h}  \right)/\sqrt{2}$ and $\theta_{\pm} = \left( \theta_{e} \pm \theta_{h}  \right)/\sqrt{2}$, and obtain a renormalized free boson model \cite{von1998,senechal2004,giamarchi2003}, which captures the low-energy physics,
\begin{align}\label{eq-boson-ham0}
	\mathcal{H}_0 = \frac{v_{\pm}}{2} \int dx\left[\frac{1}{K_{\pm}} \left( \partial_x\phi_{\pm} \right)^2 + K_{\pm} \left( \partial_x\theta_{\pm} \right)^2  \right],
\end{align}
where $ v_{\pm} = \sqrt{( v_F +(g_1\pm g_2)/2\pi ) ( v_F - (g_1\pm g_2)/2\pi 	)}$, and $K_\pm = \sqrt{(v_F - (g_1\pm g_2)/2\pi) / (v_F + (g_1\pm g_2)/2\pi) }$.
Then we include symmetry-allowed and momentum-conserved scattering processes that arise from two-body anharmonic interactions \cite{zhang2016}, which lead to the Hamiltonian $\mathcal{H}=\mathcal{H}_0+\mathcal{H}_{int}$, where
\begin{align}\label{eq-full-boson-ham}
\begin{split}
  \mathcal{H}_{int} &=  \int dx	\Big{\lbrack}  
    \alpha_1\cos(   2\sqrt{2\pi} \phi_+  )
    +\alpha_2 \cos(  2\sqrt{2\pi} \theta_- ) \Big{\rbrack},
\end{split}
\end{align}
where the first $\phi_+$-mass term is valid when a pair of ``inverted bands" are formed and the chemical potential is around the energy of band crossings;
the second $\theta_-$-mass term describes the inter-wire pair hopping process.
Notably, the $\phi_+$-mass term is absent in usual general nanowires with only electron-like (or hole-like) bands.  
In particular, the relevance of the cosine terms can be evaluated by the renormalization group (RG) equation $d\alpha_i / d\ln \lambda = [1-\Delta_{sd}(\alpha_i)]\alpha_i$, where the scaling dimensions (sd) of the coupling constants are,
\begin{align}\label{eq-scaling-dim}
\Delta_{sd}(\alpha_1) = K_+ , \quad \Delta_{sd}(\alpha_2) = \frac{1}{K_-}. 
\end{align}
Thus, under RG, we find that $\alpha_{1}$ is relevant when $K_+<1$, while $\alpha_{2}$ is relevant for $K_->1$.  
For the whole Hamiltonian $\mathcal{H}=\mathcal{H}_0+\mathcal{H}_{int}$, 
the phase diagram is determined by the Luttinger parameters $K_{\pm}$ (or equivalently $g_{1,2}$).

\begin{figure}[!htbp]
\centering
\includegraphics[width=0.8\linewidth]{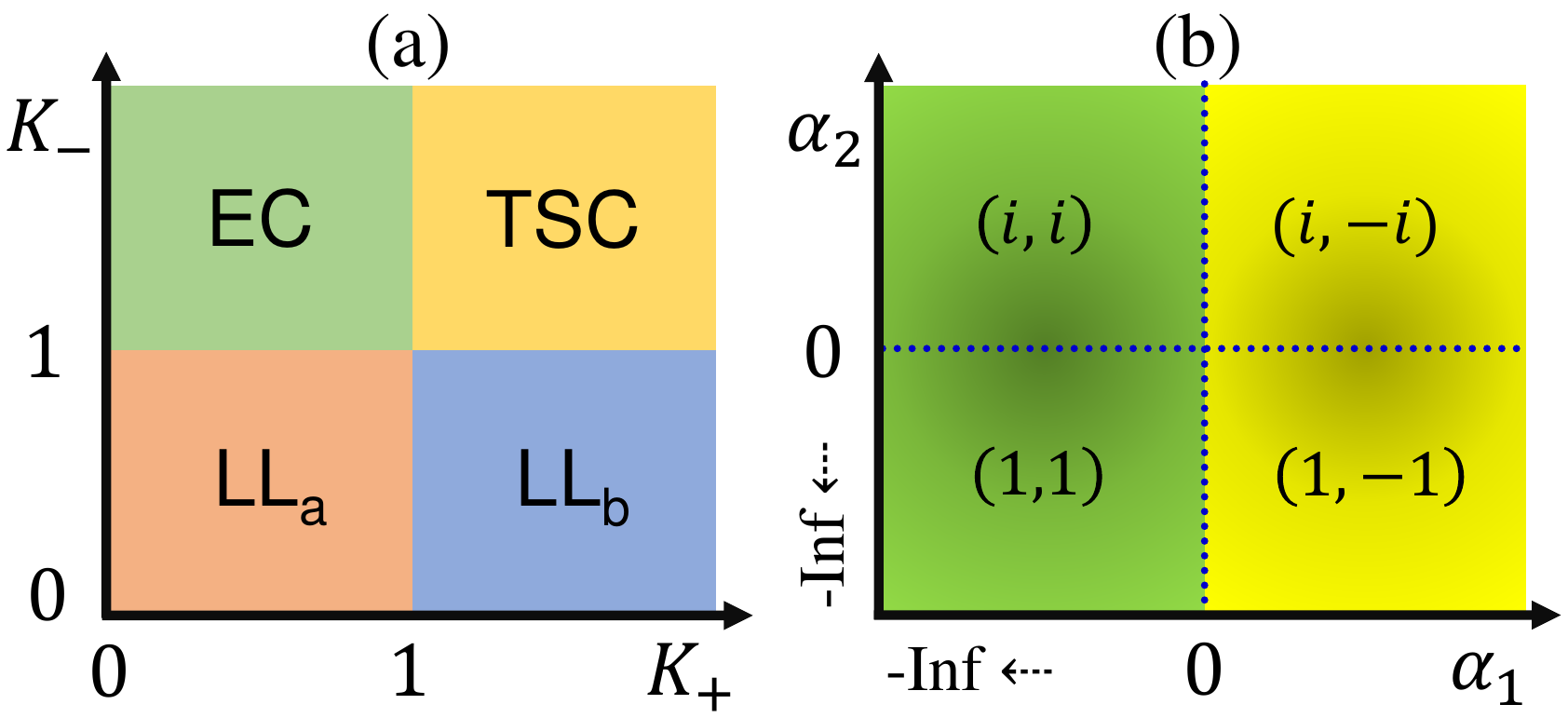}
\caption{\label{fig-phase-diagram}The phase diagrams. 
In (a), it exhibits three different phases: the exciton condensation insulating phase (EC) when $K_->1$ and $K_+<1$; the topological superconducting phase (TSC) for $K_{\pm}>1$; and the gapless Luttinger liquid phase (LL$_{a,b}$) when $K_-<1$.
(b) Given by $K_->1$ and $K_+<1$, the phase diagram for EC as functions of $\alpha_1$ and $\alpha_2$. 
When  $\alpha_1<0$, the EC phase breaks mirror symmetry $M_x$; otherwise, it is mirror symmetric. The expectation value of order parameters $(\Delta_{ex}^{(1)},\Delta_{ex}^{(2)})$ are shown in each quarter.}	
\end{figure}

We first notice that the system remains gapless when $K_+>1$. In particular, as for $K_->1$, inter-wire pair-hopping processes are greatly promoted. 
It gaps out the anti-bonding sector and further leads to number-conserving Majorana physics, which has been intensively discussed in the earlier literature \cite{fidkowski2011b,cheng2011,zhang2018b}. This phase is denoted as ``TSC" in the phase diagram in Fig. \ref{fig-phase-diagram} (a). 
On the contrary, when $K_-<1$, both anti-bonding and bonding sectors remain gapless.
This represents a gapless Luttinger liquid state [``LL$_b$" for short in Fig. \ref{fig-phase-diagram} (a)].
Similarly, a Luttinger liquid state ``LL$_a$'' arises for $K_\pm<1$ since only the bonding sector is trivially gapped. 
In this case, the anti-bonding sector could also be trivially gapped due to the pining of the $\phi_-$-mass term, 
which is induced by the Umklapp scattering when $k_F\to\pi/2$. We will discuss it later.

More importantly, we will show that the TMEI is achieved when $K_->1$ and $K_+<1$. 
Because $\Delta_{sd}(\alpha_1)<1$ and $\Delta_{sd}(\alpha_2)<1$, those cosine terms with $\alpha_{1,2}$ will flow to the strong coupling limit under RG. 
As a result, both $\theta_-$ and $\phi_+$ will be pined to the semi-classical values. 
While the system becomes gapped, we find a set of non-vanishing order parameters,
\begin{align}
\begin{split}
	&\Delta^{(1)}_{ex} \sim \left\langle  \Psi_{e,R}^\dagger \Psi_{h,L}  \right\rangle \sim \left\langle e^{-i\sqrt{2\pi}\phi_+} e^{-i\sqrt{2\pi}\theta_-}  \right\rangle \neq 0, \\
	&\Delta^{(2)}_{ex} \sim \left\langle  \Psi_{e,L}^\dagger \Psi_{h,R}  \right\rangle \sim \left\langle e^{i\sqrt{2\pi}\phi_+} e^{-i\sqrt{2\pi}\theta_-}  \right\rangle \neq 0, \\
\end{split}
\end{align}
which imply an excitonic insulating phase.
Then, let us analyze the condition for mirror-even excitonic orders.
Under $M_x$, the boson fields are transformed as: 
$ \sqrt{4\pi}\phi_e \to -\sqrt{4\pi}\phi_e + 2\beta $, $ \sqrt{4\pi}\theta_e \to \sqrt{4\pi}\theta_e + \pi$, 
$ \sqrt{4\pi}\phi_h \to -\sqrt{4\pi}\phi_h -2\beta$, and $ \sqrt{4\pi}\theta_h \to \sqrt{4\pi}\theta_h - \pi $, where $\beta$ is an unimportant phase factor. 
Consequently, the mirror symmetry sends $\phi_+\to-\phi_+$ and $\theta_-\to\theta_-+\sqrt{\pi/2}$. 
For the excitonic order parameters, we immediately find that $\Delta^{(1)}_{ex} = -\Delta^{(2)}_{ex}$ respects $M_x$, while $\Delta^{(1)}_{ex} = \Delta^{(2)}_{ex}$ does not.
The relative phase difference between $\Delta^{(1)}_{ex}$ and $\Delta^{(2)}_{ex}$ is determined by the sign of the coupling coefficients $\alpha_1$ and $\alpha_2$. 
For example, when $\alpha_{1,2}>0$, the semi-classical limit is given by  $\theta_- = (n_\theta+\frac{1}{2})\sqrt{\frac{\pi}{2}}$ and $\phi_+ = (n_\phi+\frac{1}{2})\sqrt{\frac{\pi}{2}}$. This leads to $\Delta_{ex}^{(1)} = -\Delta_{ex}^{(1)} = i$ and thus respects $M_x$ symmetry. 
In Fig. \ref{fig-phase-diagram} (b), we have mapped out the phase diagram with respect to $\alpha_1$ and $\alpha_2$. 
Therefore, the system preserves the mirror symmetry $M_x$ when $\alpha_1>0$.

To clarify the topological nature of the above mirror-even excitonic insulating physics, it is instructive to map the excitonic orders in the low-energy basis back to the original fermion basis for Eq.~\eqref{eq-ham0}. 
We find that the mirror-even case with $\Delta_{ex}^{(1)}=-\Delta_{ex}^{(2)}=1$ and $\Delta_{ex}^{(1)}=-\Delta_{ex}^{(2)}=i$ are respectively equivalent to the mean-field excitonic orders $\sigma_x\tau_y$ and $\sigma_x\tau_x$ in the original basis [see Eq.~\eqref{Eq: mirror orders}], which are already known for leading to the TMEI phase. 

Furthermore, the topological properties of the TMEI phase could also be understood via the boson theory.
Consider an open boundary condition of the system $x<0$ in Eq.~\eqref{eq-boson-ham0} with the end point $x=0$, so that $x>0$ to be the vacuum with $\phi_{\pm} = 0$ and $x<0$ to be the TMEI phase with $\phi_{+} = (n_{\phi} + \frac{1}{2})\pi/\sqrt{2\pi}$. The fractional charge bound to the system end is calculated to be 
\begin{equation}
q = e\sqrt{\frac{2}{\pi}} \int dx\partial_x \phi_+ = \left(n_\phi+ \frac{1}{2}\right)e.
\end{equation}
This half-quantized end charge thus confirms the bosonized theory with mirror-even excitonic orders as the TMEI phase. 

On the other hand, we can also explore the Luther-Emery physics with $K_+=\frac{1}{2}$, where the boson system can be reformulated into a noninteracting fermion theory. 
We focus on the bonding sector and rescale the bonding bosonic fields as $\phi_+/\sqrt{K_+}=\tilde{\phi}_+$ and $\sqrt{K_+}\theta_+=\tilde{\theta}_+$. This allows us to introduce a set of new chiral fermion operators as
\(\tilde{\psi}_{R} = \frac{\eta_{R}}{\sqrt{4\pi a}} e^{ i\sqrt{\pi}\left( \tilde{\phi}_+ + \tilde{\theta}_+ \right)} \) and \( \tilde{\psi}_{L} = \frac{\eta_{L}}{\sqrt{4\pi a}}  e^{-i\sqrt{\pi} \left( \tilde{\phi}_+ - \tilde{\theta}_+ \right)} \), which leads to the following refermionized Hamiltonian
\begin{align}\label{eq-refermionize-ham}
  \mathcal{H}_+ = \int dx\; \tilde{\psi}^\dagger \left( -iv_+ \gamma_z \partial_x + \alpha_1 \gamma_x \right) \tilde{\psi},
\end{align}
where $\tilde{\psi} =( \tilde{\psi}_{R}, \tilde{\psi}_{L})^T$ is a spinor and $\gamma_{x,y,z}$ are Pauli matrices in the new chiral fermion basis. 
Crucially, Eq.~\eqref{eq-refermionize-ham} describes the low-energy theory of a massive Dirac fermion in 1D. Since the vacuum condition pins $\phi_\pm =0$ and is equivalent to $\alpha_1<0$, the interface between the vacuum and the double wire forms a mass domain wall for the 1D Dirac fermion, which thus hosts a half-charge bound state. Therefore, Eq.~\eqref{eq-refermionize-ham} is exactly a fermionic model of TMEI phase in the strongly interacting limit, which is consistent with the above bosonic analysis.

\section{Experimental signature}
We now propose quantized transport signals in a four-terminal device to distinguish a trivial phase from the TMEI phase. 
As shown in Fig.~\ref{fig-sketch-nanowire}, the four metallic electrodes (labeled by $i=1,2,3,4$) are attached to the two-nanowire system. 
By applying a voltage drop and measuring the corresponding electric current, the conductance $G_{ij}$ between leads $i$ and $j$ can feasibly identified.
The intra-wire two-terminal conductance $G_{1,3}$ (or equivalently $G_{2,4}$) measures the electron tunneling probability across the system. Thus, $G_{1,3}$ is expected to show a U-shape dip as a function of voltage bias because of the energy gap. 

The non-local inter-wire conductance $G_{1,2}$ (or $G_{3,4}$) is the key to characterize TMEI physics. First, a non-zero $G_{1,2}$ has already implied the strong inter-wire correlation effects, which could further clarify the excitonic nature of the energy gap measured in $G_{1,3}$. For a trivial system, we thus expect a similar conductance dip for $G_{1,2}$ due to the exciton gap. For the TMEI phase, however, its half-charge end mode will provide an additional resonant inter-wire conductance contribution to $G_{1,2}$ 
\begin{equation}
	\Delta G_{1,2}=\frac{e^2}{h}\frac{\Gamma^2}{\Gamma^2 + (\omega-E_0)^2}.
\end{equation}
Here, $E_0$ is the energy of the half-charge mode and $\Gamma\sim m_0^2/v_F$ is the transport broadening. Thus, when the half-charge mode is in-gap, $G_{1,2}$ will show a quantized conductance peak of $e^2/h$ as $\omega\to E_0$. 

\begin{figure}[!htbp]
\centering
\includegraphics[width=0.9\linewidth]{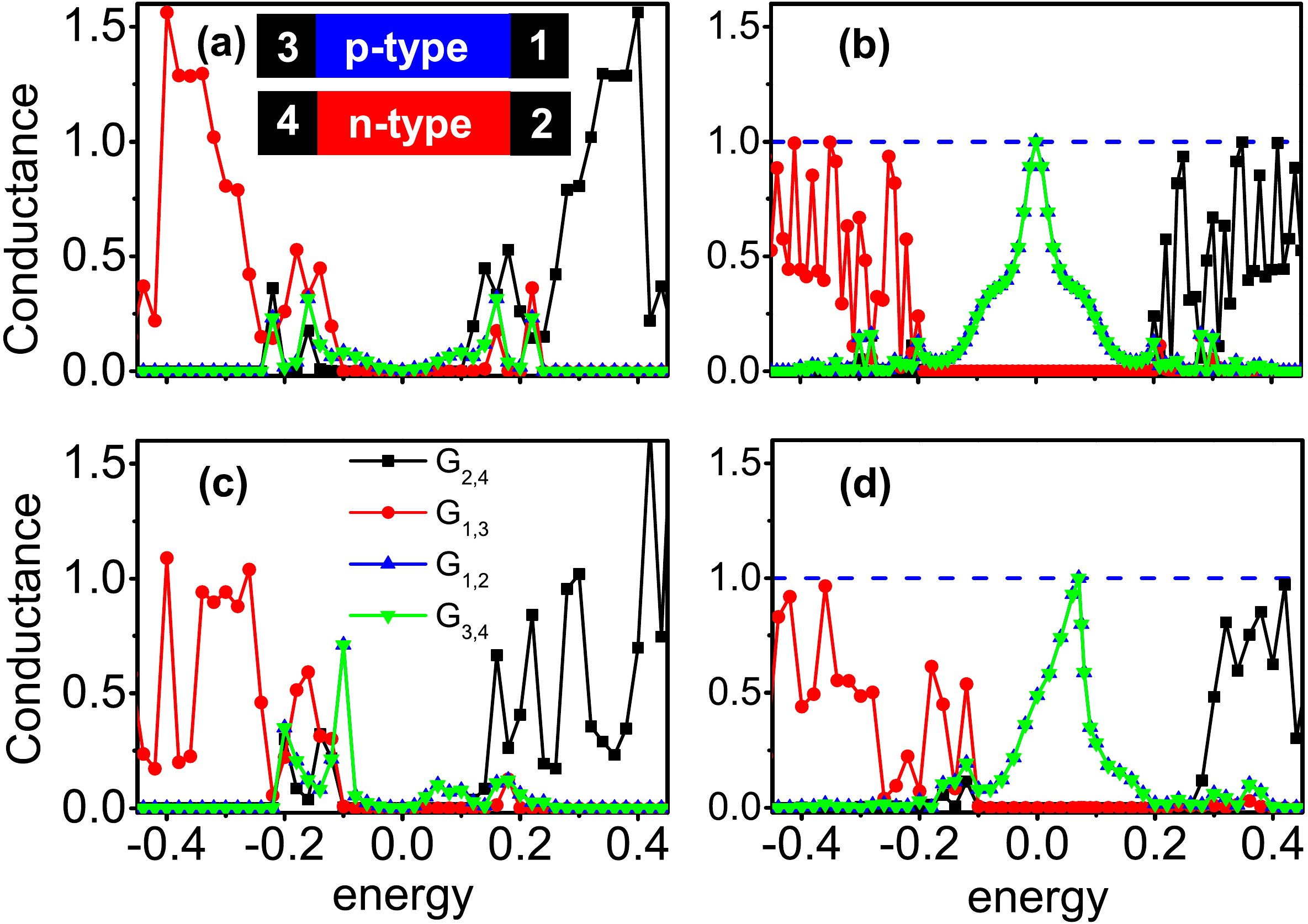}
\caption{\label{fig-transport} 
Transport conductance $G_{i,j}$ in unit of $e^2/h$ between leads. 
In (a,c) trivial and (b,d) topological phase, the bulk is fully gaped.  
Whereas the quantized $G_{1,2}$ and $G_{3,4}$ identify the half-charge end modes in the topological mirror excitonic phase.
The resonant quantization happens when energy goes to zero in (b) but finite energy in (d), due to the breaking of chiral symmetry.
Parameters used are: $m_0=1, v=1, \delta \mu=0, \mu=0, \Delta_0=0.2$; 
$\delta m=0$ in (a,b) and $\delta m=0.1$ in (c,d); 
$h=0.1$ in (a,c) and $h=0.4$ in (b,d);
The barrier potential between leads and nanowires is $V_0=1.5 m_0$	.
}	
\end{figure}

Numerical verifications of the above conductance patterns are performed using Kwant Python package \cite{groth2014}.  
As shown in Fig.~\ref{fig-transport}, [(a), (c)] and [(b), (d)] show the conductance distributions as a function of voltage bias for the topological trivial and non-trivial phase, respectively. In particular, (a) and (b) has an accidental chiral symmetry in their models, while (c) and (d) do not. 
Clearly, the conductance patterns behave exactly the same as what we predict above. While every conductance displays a finite gap for the trivial system in (a) and (c), the $G_{1,2}$ and $G_{3,4}$ for the TMEI phase show a quantized conductance peak when the energy of lead electrons matches that of the localized end states, as shown in (b) and (d). Specifically, enforcing a chiral symmetry to the system will lead to a zero-bias peak in (b). However, a general TMEI system that lacks the chiral symmetry would display a peak at finite voltage bias, as shown in (d). 
Because the evolving slightly from the free-fermion limit with $K_{\pm}=1$ to the interacting case with $K_{\pm}\neq 1$ does not close the energy gap of the system. 
Therefore, we expect that our transport results in the mean-field limit will hold for an interacting TMEI phase.

\section{Discussions and Conclusions}
Next, let us discuss the half-filling case, where one needs to consider the Umklapp scattering, 
\begin{align}\label{eq-ham-a3}
\begin{split}
H_{um} &\sim \alpha_3 \int dx\; \cos\left(  2\sqrt{2\pi} \phi_-  \right). \\
\end{split}
\end{align}
Therefore, the full boson Hamiltonian is,
\begin{align}\label{eq-full-boson-ham}
\begin{split}
\mathcal{H} &=   \frac{v_{\pm}}{2} \int dx\; \left\{  \frac{1}{K_{\pm}} \left( \partial_x\phi_{\pm} \right)^2 + K_{\pm} \left( \partial_x\theta_{\pm} \right)^2  \right\}    \\
&+	\Big{\lbrack} \alpha_1\cos\left(   2\sqrt{2\pi} \phi_+  \right)
+ \alpha_2 \cos\left(  2\sqrt{2\pi} \theta_-  \right) \\
&\quad\;+ \alpha_3\cos\left(  2\sqrt{2\pi} \phi_-  \right)  \Big{\rbrack}.
\end{split}
\end{align}
The scaling dimensions (sd) of the coupling constants $\alpha_1, \alpha_2, \alpha_3$ are,
\begin{align}\label{eq-scaling-dim}
\Delta_{sd}(\alpha_1) = K_+,\; \Delta_{sd}(\alpha_2) = \frac{1}{K_-},\; \Delta_{sd}(\alpha_3) = K_-. 
\end{align}
When $K_{\pm}<1$, we find $\Delta_{sd}(\alpha_1)<1$, $\Delta_{sd}(\alpha_2)>1$ and $\Delta_{sd}(\alpha_3)<1$. 
Namely, $\alpha_1$ and $\alpha_3$ are relevant under RG and tend to flow to the strong coupling limit. 
Consider the ``semi-classical" limit with $\alpha_{1,3}\to + \infty$, $\phi_{\pm}$ are pinned to classical values with $\phi_{\pm}=(n_\pm+1/2)\sqrt{\pi/2}$. Here $n_\pm \in \mathbb{Z}$ are integer-valued operators. 
As a result, the system develops an energy gap to all of its fermionic excitations. 
To understand the nature of this gapped phase, we define the following density-wave (DW) order parameters \cite{wu2003}:
\begin{align}
\begin{split}
&\Delta^{(1)}_{dw} \sim \langle  \Psi_{e,R}^\dagger \Psi_{e,L}  \rangle \sim \langle e^{-2i\sqrt{\pi}\phi_e} \rangle \neq 0, \\ 
&\Delta^{(2)}_{dw} \sim \langle  \Psi_{h,R}^\dagger \Psi_{h,L}  \rangle \sim \langle e^{-2i\sqrt{\pi}\phi_h}  \rangle \neq 0.
\end{split}
\end{align} 
Therefore, we notice that DW orders $\Delta_{dw}^{(1,2)}$ develop non-zero expectation values directly implies the spontaneous breaking of the translational symmetry. 
Clearly, the mirror symmetry is spontaneously broken when $\phi_-$ field is pined. 
In this case, the DW phase is trivial in terms of the mirror-protected topology, which simply because it is defined by a pinned $\phi_-$. 

Furthermore, let us briefly discuss how to realize the TMEI phase in a {\it single} nanowire of rotation-protected Dirac semimetal (e.g. a Cd$_3$As$_2$ nanowire with four-fold rotational symmetry $C_4$).
As pointed out in Ref.~\cite{zhang2018b}, applying a magnetic field along the wire will naturally drive a 1D band inversion between an electron-like band with angular momentum $J=-\frac{1}{2}$ and a hole-like band with $J=\frac{3}{2}$. 
As a result, any single-particle tunneling from the electron band to the hole band is naturally forbidden by the $C_4$ symmetry. In particular, with both $C_4$ and the spatial inversion ${\cal I}$ symmetry, the Dirac semimetal nanowire also possesses an out-of-plane mirror $M_z = C_2 {\cal I}$ that can protect the exciton-induced band topology in a corresponding nanowire geometry. Thus, without the complexity of aligning two quantum wires and careful band engineering in our double-wire setup, a single Dirac semimetal nanowire naturally fulfill all the symmetry and topological requisites for TMEI physics.

To summarize, we propose a new type of interacting crystalline topological state, the TMEI phase, that can be realized in the Rashba nanowires and Dirac semimetal nanowires. 
In particular, we have established a bosonized theory to show the robustness of TMEI phase beyond the mean-field approximation. 
This idea of exciton-induced crystalline topological states also have interesting higher-dimensional generalizations. 
For example, let us consider a bilayer two-dimensional systems with the top (bottom) layer contributing a electron (hole) band near the Fermi level. 
An out-of-plane mirror symmetry $M_z$ in this system can protect a TMEI phase with $|n_M|$ pairs of counter-propagating 1D edge modes, where $n_M\in \mathbb{Z}$ is the mirror Chern number for the system. 
On the other hand, when the bilayer system possesses in-plane mirror symmetry $M_x$ and $M_y$, it is also possible to realize a higher-order topological insulator with a quantized bulk quadruple moment and corner-localized charges. 
This is exactly an interacting and excitonic version of the electronic quadruple insulator in Ref.~\cite{benalcazar2017,peterson2018}. 
Detailed discussions on these 2d interacting crystalline topological systems will be left for future works.  
We note that the long-range excitonic orders may be realized in a 1D solid state electron systems \cite{Kantian2017}. 
It will be interesting to find the TMEI by numerical simulation, which is left for future work.

\textit{Acknowledgment.--}
R.-X.Z acknowledges Yang-Zhi Chou for helpful discussion. 
L.-H.H. and C.W. are supported by AFOSR FA9550-14-1-0168.
R.-X.Z. is supported by a JQI Postdoctoral Fellowship.
F.-C.Z. is supported by the Strategic Priority Research Program of the Chinese Academy of Sciences (Grant XDB28000000),
Beijing Municipal Science \& Technology Commission (No.Z181100004218001), 
National Science Foundation of China (Grant No.11674278), 
National Basic Research Program of China (No.2014CB921203).

\bibliographystyle{apsrev4-1}
\bibliography{ref}

\end{document}